\documentclass[apj]{emulateapj}   
\usepackage{graphicx}
\usepackage{multirow}

\usepackage{amsmath}
\usepackage{natbib}
\usepackage{hyperref}
\usepackage[usenames,dvipsnames]{color}  

\def\msun{{\rm ~M}_{\odot}}
\def\rsun{{\rm ~R}_{\odot}}

\begin{document}

\title{The Formation of a $70\msun$ Black Hole at High Metallicity}

\author{
   K.Belczynski\altaffilmark{1}, R.Hirschi\altaffilmark{2,3}, E.A.Kaiser\altaffilmark{3},
   Jifeng Liu\altaffilmark{4,11}, J.Casares\altaffilmark{5,6}, Youjun Lu\altaffilmark{4,11},     
   R.~O'Shaughnessy\altaffilmark{7}, A.Heger\altaffilmark{8,9}, S.Justham\altaffilmark{10,11},  
   R.Soria\altaffilmark{4,11}
} 

\affil{
   $^{1}$ Nicolaus Copernicus Astronomical Center, Polish Academy of Sciences,
          ul. Bartycka 18, 00-716 Warsaw, Poland\\ (chrisbelczynski@gmail.com)\\  
   $^{2}$ Kavli IPMU (WPI), The University of Tokyo, Kashiwa, Chiba 277-8583, Japan \\
   $^{3}$ Astrophysics group, Keele University, Keele ST5 5BG, United Kingdom\\
   $^{4}$ Key Laboratory of Optical Astronomy, National Astronomical Observatories, 
          Chinese Academy of Sciences, Beijing 100101, China\\
   $^{5}$ Instituto de Astrof\'\i{}sica de Canarias, c/Via Laceta s/n, E-38205 La 
          Laguna, Tenerife, Spain\\
   $^{6}$ Departamento de Astrof\'\i{}sica, Universidad de La Laguna, E-38206 La Laguna, 
          Tenerife, Spain \\
   $^{7}$ Rochester Institute of Technology, Rochester, New York 14623, USA\\
   $^{8}$ School of Physics and Astronomy, Monash University, Victoria 3800, Australia\\
   $^{9}$ OzGrav: Australian Research Council Centre of Excellence for Gravitational Wave 
          Discovery, Clayton, VIC 3800, Australia\\
   $^{10}$ Anton Pannekoek Institute of Astronomy and GRAPPA, University of
           Amsterdam, 1090 GE Amsterdam, the Netherlands\\
   $^{11}$ School of Astronomy and Space Science, University of the Chinese
           Academy of Sciences, Beijing 100012, China
}
\begin{abstract}
 
A $70\msun$ black hole was discovered in Milky Way disk in a long period ($P=78.9$ 
days) and almost circular ($e=0.03$) detached binary system (LB-1) with a high 
($Z \sim 0.02$) metallicity $8\msun$ B star companion. 
Current consensus on the formation of black holes from high metallicity stars limits 
the black hole mass to be below $20\msun$ due to strong mass loss in stellar winds. 
So far this was supported by the population of Galactic black hole X-ray binaries 
with Cyg X-1 hosting the most massive $\sim 15\msun$ black hole.
Using the Hurley et al. 2000 analytic evolutionary formulae, we show that the formation 
of a $70\msun$ black hole in high metallicity environment is possible if stellar wind 
mass loss rates, that are typically adopted in evolutionary calculations, are reduced 
by factor of $5$. As observations indicate, a fraction of massive stars ($\sim 7\%$) 
have surface magnetic fields which, as suggested by Owocki et al. 2016, may quench the 
wind mass-loss, independently of stellar mass and metallicity. 
We also computed detailed stellar evolution models and we confirm such a scenario. 
A non-rotating $85\msun$ star model at $Z=0.014$ with decreased winds ends up as a 
$71\msun$ star prior core-collapse with a $32\msun$ helium core and a $28\msun$ CO core. 
Such star avoids pair-instability pulsation supernova mass loss that severely limits 
black hole mass and may form a $\sim 70\msun$ black hole in the direct collapse. 
Stars that can form $70\msun$ black holes at high $Z$ expand to significant size with 
radius of $R \gtrsim 600 \rsun$ (thanks to large H-rich envelope), however, exceeding 
the size of LB-1 orbit (semi-major axis $a\lesssim350\rsun$). Therefore, we can 
explain the formation of black holes upto $70\msun$ at high metallicity and this result 
is independent from LB-1; whether it hosts or does not host a massive black hole.  
However, if LB-1 hosts a massive black hole we are unable to explain how such a binary 
star system could have formed without invoking some exotic scenarios.

\end{abstract}


\section{Introduction}
\label{sec.intro}

LB-1 is reported as a detached binary system containing B star with a mass of $8\msun$
($-1.2/+0.9\msun$) and a black hole (BH) with a mass of $68\msun$ ($-13/+11\msun$).
The binary system orbit is almost circular with $e=0.03$ ($-0.01/+0.01\msun$) and has 
an orbital period of $P_{\rm orb}=78.9$ days ($-0.3/+0.3$ days). This corresponds to a
physical semi-major axis of $a=300-350\rsun$ and a Roche lobe radius of the BH 
$R_{\rm BH,lobe} \lesssim 200\rsun$. This system is one of the widest known binary 
system hosting a stellar-origin BH, see \url{https://stellarcollapse.org}.
Two other binaries, proposed to host BH candidates, were also discovered by the radial 
velocity method by \cite{Thompson2019} and \citet[although this is in a globular 
cluster and has a very large period, $P=167$\,d, and eccentric  orbit with $e=0.6$ and 
it must have formed by capture]{Giesers2018}. 

The LB-1 was discovered by the 4-meter class telescope LAMOST and the spectroscopic 
orbit was confirmed by the 10-meter class Gran Telescopio Canarias and Keck telescopes. 
Chandra non-detection places X-ray emission at the very low level $<2 \times 10^{31}$ 
erg/s. An $H_\alpha$ emission line was observed, however, and since it follows a 
BH (small accretion disk around the BH from the B star wind) the double spectroscopic 
orbital solution was obtained. The system is on the outskirts of the Galactic 
disk, in the anti-Galactic center direction, about $4$\,kpc away from Sun. There is no 
globular cluster nearby ($<4$kpc). The chemical composition of B star indicates 
a slightly over-Solar metal abundance $Z=0.02$ assuming $Z_\odot = 0.017$. The full 
information on the system parameters and the discovery is reported in~\cite{Liu2019}.  

Since the publication of the discovery paper, there are a number of studies that 
attempt either to reject specific formation scenarios of LB-1 \citep[the massive BH 
is the BH-BH merger product or a very close BH-BH binary; see][]{Shen2019} or to 
explain it with some specific scenarios: stellar evolution of a massive magnetic star
~\citep{Groth2019}, merger of two unevolved stars~\citep{Tanikawa2019}, merger of a 
BH and an unevolved star~\citep{Banerjee2019,Olejak2019}. It was also pointed out 
that the existence of LB-1 (and its future evolution) may be in tension with the 
non-detection of ultra-luminous X-ray sources or black hole neutron star systems in 
the Galaxy~\citep{Safarzadeh2019}. Alternatively, the nature of LB-1 is questioned 
with a reanalysis of observational data and results that support the idea that either 
the BH or both components are of lower mass than originally claimed
~\citep{AbdulMasih2019,ElBadry2019,Eldridge2019,SimonDiaz2019,Irrgang2019}. 
This would allow the classical isolated binary evolution at high metallicity to 
explain the formation of LB-1.

In fact, the existence of a $70\msun$ BH in high metallicity environment seems 
challenging. The current consensus is based on mass loss rate estimates and their 
dependence on metallicity for H-rich stars~\citep{Vink2001} and He-rich stars
~\citep{Vink2005,Sander2019} that seems to limit BH mass to about $20\msun$ at solar 
metallicity~\citep{Belczynski2010b}. Existing electromagnetic observations seem to 
support this paradigm \citep{Casares2014}. Note the masses of the two most massive 
stellar-origin BHs that are known to have formed at relatively high metallicity are 
the well known Cyg X-1~\citep[$M_{\rm BH}=14.8 \pm 1.0$ $Z \approx 0.02$;][]{Orosz2011} 
and M33 X-7~\citep[$M_{\rm BH}=15.7 \pm 1.5$, $Z \approx 0.1\,Z_\odot$;][]{Valsecchi2010}. 

The mass of the BH in LB-1 seems to contradict pair-instability pulsation supernovae 
(PPSN) and pair-instability supernova (PSN) theory, that limits BH mass to about 
$M_{\rm BH}<40-50\msun$ \citep{Bond1984a,Heger2002,Woosley2017,Farmer2019,Leung2019}.
This limit was recently proposed to be as high as $\sim 55\msun$ for non-zero 
metallicity stars (Population I/II) by~\cite{Belczynski2017b}. Note that for the 
LIGO/Virgo most massive BH-BH merger in O1/O2 (GW170729), the primary BH mass was reported 
to be $51.2\msun$. This high mass (not the merger itself) is likely to be a statistical 
fluctuation~\citep{Fishbach2019}. However, even such mass can be explained as long as
the BH was formed at low metallicity. 
The PPSN/PSN instability can be avoided (at best) for a $70\msun$ star that can 
possibly produce $69\msun$ BH if only small neutrino mass loss takes place at the 
BH formation. This was envisioned for an ultra-low metallicity and Population III 
stars as they can keep massive H-rich envelopes \citep{Heger2002,Woosley2017}. 

Here, we propose that a similar mechanism may also work at high metallicity. The 
modification that we need to introduce to stellar evolution is to lower wind mass 
loss rates for (at least some) massive stars. The empiric diagnostics of winds of 
the massive stars are complex, especially because of the wind clumping
~\citep{Fullerton2006,Oskinova2007} and the agreement between theory and 
observations are not always conclusive~\citep{Keszthelyi2017}.

In lower metallicity environments, such as in the LMC and the SMC galaxies, some 
work~\citep{Massa2017} indicates that wind mass loss rates may be actually higher 
than typically adopted in evolutionary predictions~\citep{Vink2001,Belczynski2010b}, 
others seem to agree with standard calculations~\citep{RamirezAgudelo2017}, and yet 
others point out to much lower mass-loss rates than expected~\citep{Bouret2003,
Ramachandran2019,Sundqvist2019}. In the upper stellar mass regime, \cite{Hainich2013,
Hainich2019} determine mass-loss rates which are in broad agreement with the 
theoretical expectations.

In this work we consider the mass regime $70-100\msun$ at solar metallicity. 
\cite{Vink2012} have shown that for stars in transitional regime (from optically 
thin to thick winds) the standard mass loss rates should apply. The empirical studies 
that include hydrogen-rich Wolf-Rayet stars~\citep{Hamann2019} find mass-loss rates 
lower than theoretically predicted~\cite{Nugis2000} for the most luminous objects. 
However, what are the mass-loss rates of such massive stars when they are very 
young is not well known. \cite{Gruner2019} found that the mass-loss rate of the 
earliest O-type star in the Galaxy (HD 93129A, the primary mass is $\sim 100\msun$) 
compares well with the theoretical expectations, but this result depends on assumed 
clumping parameters. Furthermore, about $7\%$ of OB stars are known to have (mostly) 
dipolar magnetic fields~\citep{Fossati2015,Wade2016,Grunhut2017}. Some of these known 
magnetic stars are massive, but not quite reaching the mass regime considered here 
unless errors on mass estimates are considered: $61\pm33\msun$ for CPD-28 2561
or $\sim 60\msun$ for HD 148937~\citep{DavidUraz2019}. These magnetic fields may 
capture wind particles and reduce wind mass loss rates independent of star mass and 
metallicity~\citep{Owocki2016,Petit2017,Shenar2017,Georgy2017}. Here we show two 
things. First, that the decrease of wind mass loss rates (independent of the reduction 
origin) allows some models to avoid pair-instability associated mass loss and allow for 
the formation of high mass BHs ($\sim 50-70\msun$) at high metallicity. Second, that we 
are not able to make such a massive BH progenitor star fit within binary orbit of LB-1, 
if in fact LB-1 hosts a $70\msun$ BH.

\section{Calculations}
\label{sec.calc}
 
\subsection{Simple {\tt StarTrack} Simulation}
\label{sec.calc1}

\begin{figure}
\hspace*{-0.4cm}
\includegraphics[width=0.5\textwidth]{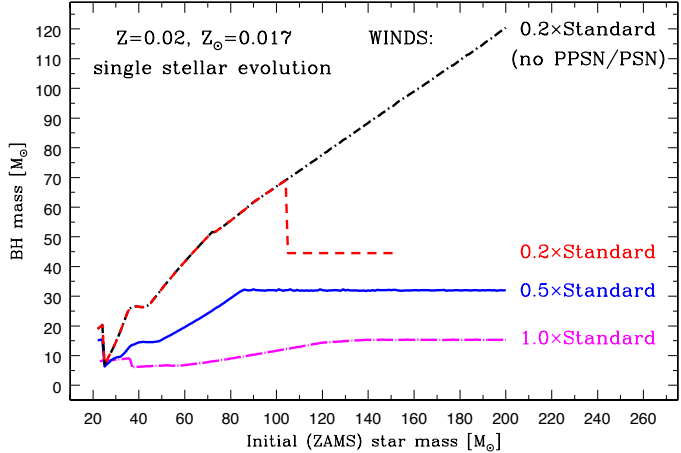}
\caption{ 
Black hole mass for single stars at metallicity estimated for LB-1 as a function 
of initial star mass. For standard wind mass loss prescriptions only low-mass 
black holes are predicted: $M_{\rm BH}<15\msun$. For reduced wind mass loss, 
however, much heavier black holes are formed: $M_{\rm BH}=30\msun$ for winds 
reduced by factor of $2$, and $M_{\rm BH}=70\msun$ for winds reduced by factor 
of $5$ of the standard values. Note that to reach even higher masses it is 
needed to switch off pair-instability pulsation supernovae that severely limit 
black hole masses.
}
\label{fig.bhmass}
\end{figure}

We used the population synthesis code {\tt StarTrack}~\citep{Belczynski2002,
Belczynski2008a} to quickly test the possibility of the formation of a $70\msun$ 
BH with decreased wind mass loss. We employed the rapid core-collapse supernova 
(SN) engine NS/BH mass calculation ~\citep{Fryer2012}, with strong PPSN/PSN 
mass loss~\citep{Belczynski2016c}. Standard winds for massive stars are used
as the base model: O/B star \cite{Vink2001} winds and LBV winds \citep[specific 
prescriptions for these winds are listed in Sec.2.2 of][]{Belczynski2010b}. 
In wind mass loss prescriptions we introduce a multiplication factor that for 
our standard calculation is $f_{\rm wind}=1.0$. Note that such approach produces 
a maximum of $\sim 15\msun$ for BHs at high metallicity ($Z=0.02$ assuming 
$Z_\odot = 0.017$) as demonstrated in Figure~\ref{fig.bhmass}. 
We also calculate evolution of single stars for decreased winds for two extra 
models with $f_{\rm wind}=0.5, 0.2$. It is clear from Figure~\ref{fig.bhmass} 
that winds need to be reduced by a factor of  $\sim 5$ to produce a $\sim 70\msun$ 
BH at high metallicity. 

Our specific example is a star with $M_{\rm zams}=104\msun$ at $Z=0.02$ and
the star is evolved with \cite{Hurley2000} analytic formulae (used in many
population synthesis and globular cluster evolutionary codes). H-rich wind
mass loss rates are decreased with $f_{\rm wind}=0.2$. The star keeps its H-rich
envelope throughout the entire evolution. After $3.8$ Myr of evolution, the star 
has a mass of $M_{\rm tot}=69.8\msun$ with a H-rich envelope mass of 
$M_{\rm env}=24.8\msun$, He core mass of $M_{\rm He}=44.99\msun$, and CO
core mass of $M_{\rm CO}=34.8\msun$. According to the simplistic population 
synthesis prescription \citep[no PPSN/PSN for stars with 
$M_{\rm He}<45.0\msun$;][]{Woosley2017} this star is not yet the subject to 
PPSN/PSN. The star undergoes core-collapse and with $1\%$ neutrino mass loss 
it forms a BH through direct collapse: $M_{\rm BH}=69.1\msun$. 

\subsection{Single Star Evolutionary Models}
\label{sec.calc2}

To explore the possibility of the LB-1 black hole being the descendant of 
a single star and to test the simple estimates from Section~\ref{sec.calc1}, 
we ran a series of stellar evolution models using the {\tt MESA} code revision 
11701~\citep{Paxton2015}. We used a solar initial composition of $Z=0.014$ for 
all models with an \cite{Asplund2009} metal mixture ({\tt initial\_zfracs = 6}), 
and the corresponding opacity tables  ({\tt kappa\_file\_prefix = `a09'}) 
including low-temperature tables ({\tt kappa\_lowT\_prefix = `lowT\_fa05\_a09p'}) 
and C/O-enhanced (type 2) opacity tables ({\tt kappa\_CO\_prefix = `a09\_co'}). 
For convection, we used the Schwarzschild boundary location condition and 
included convective boundary mixing with a value of the exponentially-decaying 
diffusion coefficient parameter $f$ and $f_0$ everywhere equal to $0.004$. 
For the reaction network, we used the {\tt basic.net} and 
{\tt auto\_extend\_net = .true.}, with which {\tt MESA} adapts the network 
along the evolution to the smallest network needed to trace energy generation. 
The main ``stabilizing'' setting/approximation was the use of extra pressure 
at the surface of the star by setting {\tt Pextra\_factor = 2}. Another one 
was the use  of {\tt MLT++} \citep[see][Sect. 7]{Paxton2013}. These settings 
might underestimate the radius of the star in our models. The models were 
evolved at least until the end of core He-burning and generally stopped due 
to convergence issues near the end of core carbon burning.

We used the ``Dutch'' scheme for mass loss with a default 
{\tt Dutch\_scaling\_factor = 1.0}. The two main mass loss prescriptions 
experienced by our hydrogen-rich models are \cite{Vink2001} for hot stars and 
\cite{deJager1988}, which we used for the cool ``Dutch'' wind. In order to reduce 
the mass loss rates, we lowered the {\tt Dutch\_scaling\_factor} by introducing a 
multiplication factor in front of wind mass loss rates and we changed it in wide 
range $f_{\rm wind}=1.0-0.0$. We calculated non-rotating and
rotating models (see Tab.~\ref{tab.mesa}). The standard rotation settings were 
used ~\citep{Heger2000}. Rotation is set on the zero-age main-sequence and the 
initial rotation rate, in terms of $\Omega$/$\Omega_{\rm crit}$, is given in 
Table~\ref{tab.mesa}. We include the following rotation-induced instabilities; 
Eddington-Sweet circulation, secular shear instability and Taylor-Spruit dynamo 
~\citep{Spruit2002}. Table~\ref{tab.mesa} gives key properties of representative 
stellar models. Using the physical ingredients described above and considering 
that the main uncertainty in the models is mass loss, we reduced the mass loss 
with a multiplication factor given in the Table in an attempt to produce a final 
total mass equal to that of LB-1, i.\,e. around $70\msun$.

\begin{table}
\caption{Initial mass, rotation and mass loss re-scaling factor (columns 1-3) 
and final total, He- and CO-cores masses and maximum radius (columns 4-7) of 
the stellar models.}
\begin{tabular}{c c c | c c c c }
\hline
$M_{\rm zams}$ & $\Omega/\Omega_{\rm{crit}}$ & $f_{\rm wind}$ & $M_{\rm{tot}}$ & $M_{\rm{He}}$ & $M_{\rm{CO}}$ & $R_{\rm max}/\rsun$ \\
\hline
\multicolumn{7}{c}{Non-rotating models}\\
\hline
100 & 0.0 & 0.576 & 70.8 & 41.5 & 36.9 & 711.1 \\
85 & 0.0 & 0.333 & 70.9 & 31.6 & 27.6 & 653.9 \\
70 & 0.0 & 0.0 & 70.0 & 30.8 & 27.0 & 637.5 \\
\hline
\multicolumn{7}{c}{Rotating models} \\
\hline 
100 & 0.6 & 0.576 & 61.6 & 49.5 & 43.9 & 260.8 \\
85 & 0.6 & 0.576 & 58.2 & 40.3 & 35.4  & 363.9 \\
85 & 0.6 & 0.333 & 62.9 & 46.8 & 41.3 & 235.0 \\
75 & 0.6 & 0.576 & 53.9 & 34.5 & 30.1  & 376.5 \\
70 & 0.6 & 0.576 & 50.2 & 32.1 & 27.8  & 324.1 \\
70 & 0.4 & 0.282 & 58.5 & 32.5 & 28.3  & 611.8 \\
\hline
\multicolumn{7}{c}{Rotating models losing entire H-layer} \\
\hline 
100 & 0.6 & 1.0 & 40.5 & 40.5 & 36.8 & 170.9 \\
100 & 0.8 & 0.882 & 43.4 & 43.4 & 37.5 & 165.5 \\
\hline
\hline
\end{tabular}
\label{tab.mesa}
\end{table}

\begin{figure}
\hspace*{0.0cm}
\includegraphics[width=0.5\textwidth]{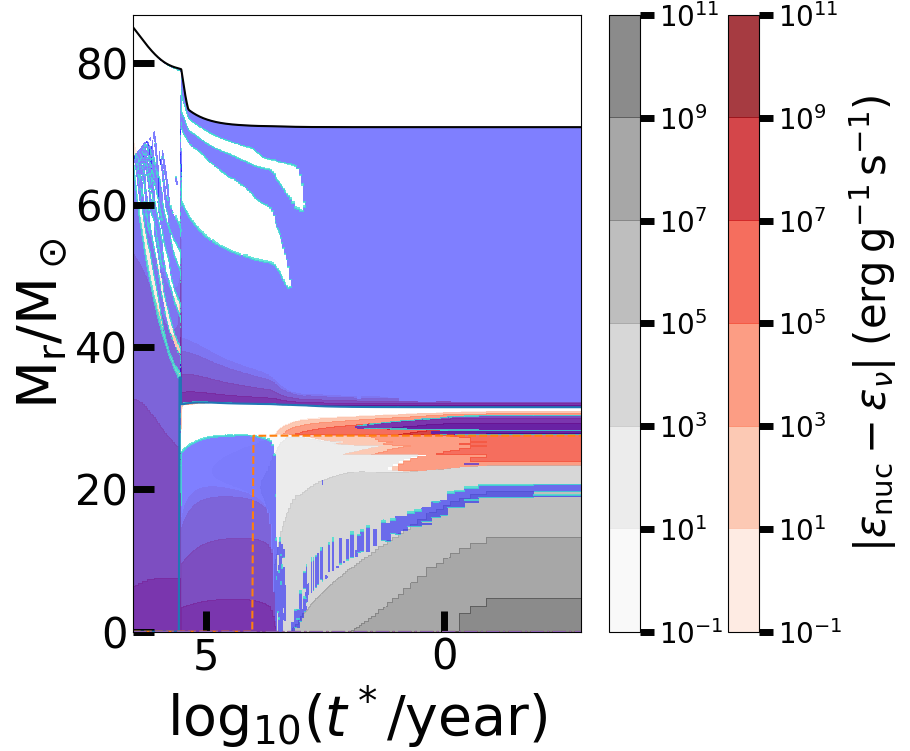}
\caption{ 
Stellar evolution diagram of the $M_{\rm zams}=85\msun$ non-rotating model with 
low stellar winds (reduced by a factor of $3$ compared to the default) at 
$Z=0.014$. The blue regions show the convective regions. Red shading indicates 
nuclear energy generation and grey shading indicates regions where cooling by 
neutrino emission dominates. The evolution of the star is presented as a 
function $t^*$, the time left until collapse/last model. The diagram presents the 
end of core hydrogen burning (left side), core helium burning and carbon burning 
(purple right). The top black solid lines indicates the total mass and the red 
dashed line indicates the He-free/poor core (defined as region where mass 
fraction of He is less than one percent). This model produces a $70.9\msun$ star 
at core collapse with a He core of $M_{\rm He}=31.6\msun$ and CO core of 
$M_{\rm CO}=27.6\msun$ and is most likely not subject to pair-instability 
pulsation supernova mass loss. This model can thus form a $70\msun$ black hole 
if there is no mass loss at BH formation. 
}
\label{fig.kip85}
\end{figure}

\begin{figure}
\hspace*{0.0cm}
\includegraphics[width=0.5\textwidth]{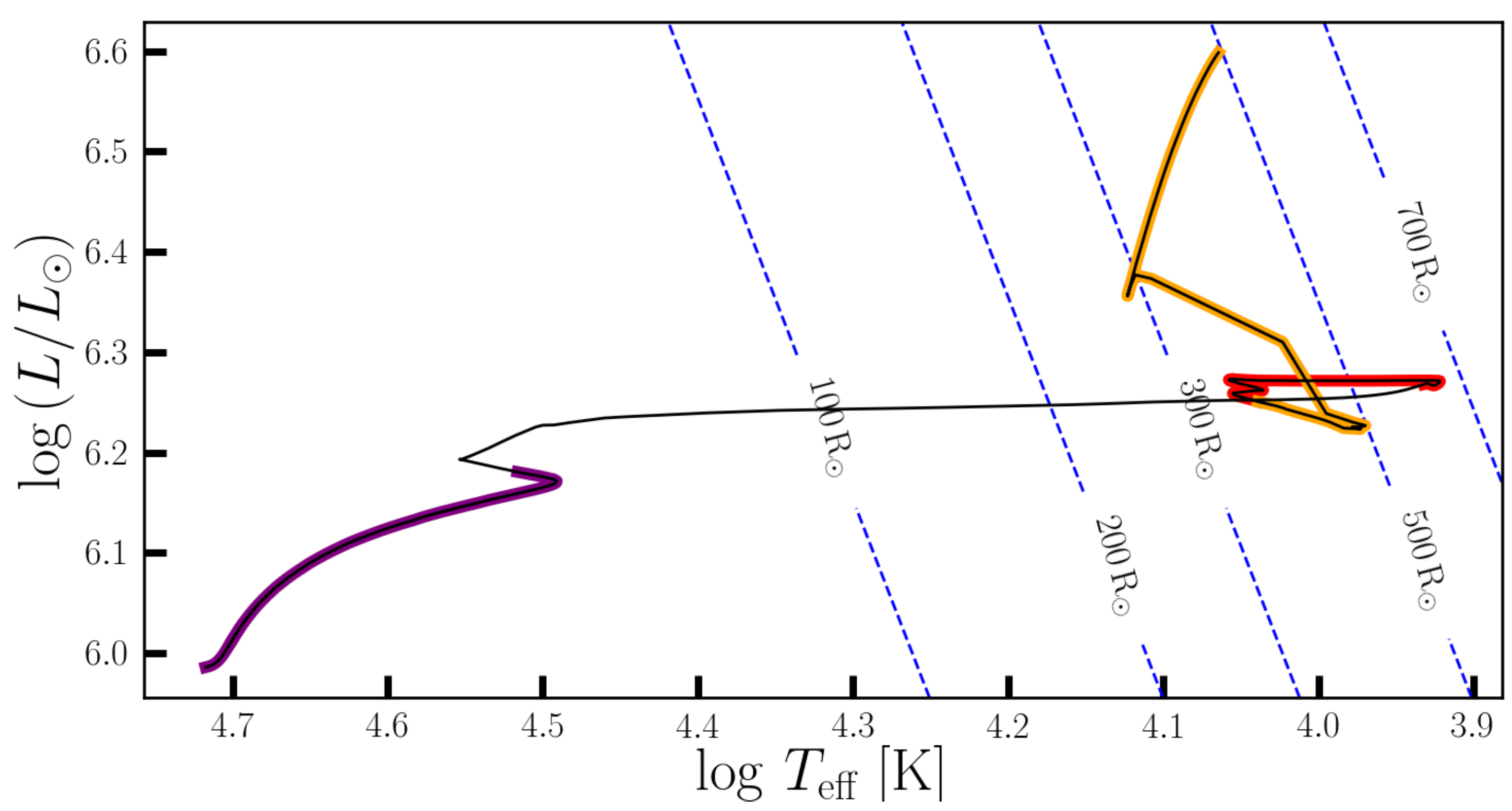}
\caption{ 
Hertzsprung-Russell diagram of the $M_{\rm zams}=85\msun$ non-rotating model 
with reduced stellar winds (by a factor of $3$ compared to default settings). 
The central burning phases are highlighted, with purple for hydrogen, red for 
helium and orange for carbon burning. The blue dashed lines indicate contours 
of constant radii. This model expands to a maximum radius of 
$R_{\rm max}\approx650\rsun$ before it loses mass during He-burning.
}
\label{fig.hr85}
\end{figure}

Considering first non-rotating models, a model without mass loss 
($M_{\rm zams}=70\msun$, $f_{\rm wind}=0.0$) is also included for reference 
as the most extreme (and unrealistic) case. With th re-scaled wind by  
$f_{\rm wind}=0.576$, a model with an initial mass of $100\msun$ ends with a 
total mass $70.8\msun$. This model has final core masses that will experience 
pair-instability pulsation mass loss, however, and thus lose more mass before 
it produces a BH. Furthermore, its radius is too large to fit in the orbit of 
LB-1. The most interesting model is the $M_{\rm zams}=85\msun$ with 
$f_{\rm wind}=0.333$. The final total mass is $70.9\msun$ and very importantly 
the final CO core mass is below the limit for pair-instability pulsation 
supernova mass loss. Indeed, the CO core mass of this model is $M_{\rm CO}=27.6\msun$ 
(see Fig.~\ref{fig.kip85}), which is below the CO core mass threshold for PPSN 
according to Table 1 in \citet[no pulsations for models with CO core masses 
below $28\msun$]{Woosley2017}. It is thus possible for this model to produce a 
$70\msun$ BH. Unfortunately, the maximum radius of this model 
($R_{\rm max}\approx650\rsun$; see Fig.~\ref{fig.hr85}) is too large to fit in 
Roche lobe of the LB-1's BH ($<200\rsun$) and this model thus cannot provide a 
full solution for the origin of LB-1. 

Considering rotating models, rotation-induced mixing leads to more massive cores 
and more mass loss~\citep[e.g.,][]{Hirschi2004}. Thus the rotating models with 
similar initial parameters end with smaller total masses and larger core masses, 
which makes them less suitable candidates to explain LB-1. The only advantage of 
rotating models over non-rotating ones is that they end with smaller radii that 
could possibly fit in the LB-1. So we may then ask the question: what is the most 
massive final single star model that would always fit in LB-1? Considering models 
that lose the entire H-rich layers (e.g. $100\msun$ models at the bottom of Tab.
~\ref{tab.mesa}) or pure He-star models of~\cite{Woosley2017} or ~\cite{Farmer2019}, 
BH masses up to $45-50\msun$ can be produced. Since He-stars are very compact, 
these would fit within the LB-1 BH Roche lobe but the BH mass would be below the 
current lower mass limit of $55\msun$ for LB-1. We also consider rotating models 
with $f_{\rm wind}=0.576$ that do not lose H-rich layers: 
$M_{\rm zams}=70, 75, 85, 100\msun$. The $70\msun$ model has a final CO 
core mass below the pair-instability pulsation mass range so is likely to collapse 
to a BH with little mass loss. The final radius, however is not so small 
($R_{\rm max}=324\rsun$) and that model would not fit in Roche lobe of LB-1's 
primary and the total mass is smaller than the lower mass estimate of BH mass in 
LB-1. The $75$ and $85\msun$ have a larger final masses but also larger CO core 
masses and radii so will likely lose some mass by pair-instability supernova 
pulsations and would not fit in LB-1. The $100\msun$ model produces relatively 
small maximum radius ($R_{\rm max}=261\rsun$) but still it would not fit in Roche 
lobe of LB-1's primary. Although final model mass is large (above lower limit on 
LB-1 BH mass), this model has a massive CO core and is subject to strong pair-instability 
pulsation supernova mass loss. A similar case is found for $85\msun$ rotating model 
with $f_{\rm wind}=0.333$. It thus seems very unlikely for a single star or a 
non-interacting star in a binary system to produce the BH in LB-1 with the currently 
derived properties.

\section{Discussion and Conclusions} 
\label{sec:concl}

It is generally believed that Population I/II stars cannot form BHs in the mass 
range $\sim 55-135\msun$, the so-called the second mass gap, due to mass loss in 
pair-instability pulsation supernovae and due to total star disruption by 
pair-instability supernovae. It was noted, however, that in one specific case the 
lower bound of the second mass gap can be shifted to $\sim 70\msun$. Such case 
was proposed for metal-poor (Population III) stars, for which wind mass loss is 
negligible even for high mass stars and then such stars can retain massive H-rich 
envelopes throughout their evolution. Retention of massive H-rich envelope allows 
a star to ignite a H-burning shell, which supports the outer stellar layers and 
helps density/temperature in stellar interior to avoid pair-instability regime 
(where adiabatic index becomes small $\gamma<4/3$). In principle, one can imagine 
a stable (against PPSN/PSN) stellar configuration with $70\msun$ star at the 
core-collapse, with He core mass of $\lesssim 40\msun$ and H-rich envelope of 
$\gtrsim 30\msun$ for a metal-poor star (for which mass loss is expected to be 
low, at least lower than at high metallicity).  

We found that similar configuration can be achieved for high metallicity stars
if wind mass loss rates are decreased in stellar evolution models. For one model, 
a non-rotating $M_{\rm zams}=85\msun$ and $Z=0.014$ star, we can form a $70\msun$ 
BH as a single star or a binary component in a very wide non-interacting binary if 
standard wind mass loss rates are reduced by factor of $\sim 5$. This is rather 
surprising and unexpected result on its own. Note that this result is totally 
independent of LB-1 and its true nature; whether it hosts a massive BH or not. 
This model, however, is not useful in context of LB-1 as the stellar radius of 
this star ($\gtrsim 650\rsun$) is too large to fit within LB-1 orbit. 

The main uncertainty in the massive star models is mass loss. We reduced the mass 
loss rates in order to produce higher final masses. Note that reduced wind mass 
loss does not have to operate for all stars, but maybe it is possible that wind is
quenched only for some fraction of very massive stars (e.g., via magnetic capture 
of wind particles: see Sec.~\ref{sec.intro}). Other studies 
\citep[e.g.,][]{Limongi2018,Chieffi2019}, however, show that a higher mass loss is 
needed in the red supergiant (RSG) phase to reproduce the absence of observed type 
II SNe above a certain luminosity \citep{Smartt2009}. Evolved massive stars are also 
expected to lose mass via eruptive events, e.g. LBV-type mass loss, beyond the 
Humphreys-Davidson limit \citep{Humphreys1979,Langer2012,Smith2014}. These extra 
mass loss was suggested to explain the apparent lack of cool luminous massive stars 
in Milky Way~\citep{Mennekens2014}. Note that our model of non-rotating $85\msun$ star 
that can produce $70\msun$ BH enters cool (${\rm log_{10}}(T_{\rm eff}) \approx 3.9$) 
and luminous (${\rm log_{10}}(L/L_\odot) \approx 6.3$) region part of H-R diagram 
(see. Fig.~\ref{fig.hr85}). Even at low metallicity of Small Magellanic Cloud stars 
are not found at such low temperatures and such high luminosities (see Fig.13 of 
~\cite{Ramachandran2019}).

Therefore, the existence of LB-1, if it really hosts a massive $70\msun$ BH, 
points to some other possibilities. 
{\em (i)} Either pair-instability does not operate in stars as expected. This would 
allow a rapidly rotating massive star to evolve homogeneously keeping small radius 
and forming $70\msun$ helium-rich object that would directly collapse to 
a black hole. 
{\em (ii)} Or the BH is a descendant of BH-BH or BH-star merger in inner binary and 
LB-1 was originally a triple system. Note that this would also require homogeneous 
evolution of two $\sim 30-50\msun$ stars as not to affect nearby B star, but this 
would not require violating pair-instability theory. However, a gravitational-wave 
kick during BH-BH merger or any natal kick at BH formation, may be incompatible with 
very low eccentricity of LB-1.  
{\em (iii)} Maybe some stars expand less due to exotic composition and modifications 
of opacities or to an unknown additional mixing process.
Alternatively, LB-1 may have lower mass components than claimed in the discovery 
paper and then standard stellar/binary evolution can account for the formation of 
such system. 

Note that if BHs as massive as $70\msun$ exist in young and metal-rich environments, 
e.g., Galactic disk, they would most likely have low spins since our models employ 
effective angular momentum transport by magnetic dynamo 
\citep[$a \lesssim 0.15$; see][]{Belczynski2017b}. If such massive BH could catch a 
companion, e.g., in an open cluster, or have formed in a wide binary with another  
BH that then evolves into close/merging system,  e.g., by a ``lucky'' natal kick 
injection into short period and eccentric orbit, then LIGO/Virgo will sooner or 
later discover these massive BHs. LIGO/Virgo detection of objects of such mass will 
be burdened with large errors, $\sim 20-30\msun$ up and down, so in principle even 
a detection of a $100\msun$ BH could be possibly explained by one of our models.

\acknowledgements
Authors would like to thank Lida Oskinova and anonymous reviewer for useful
comments. 
KB acknowledges support from the Polish National Science Center (NCN) grant
Maestro (2018/30/A/ST9/00050). KB and RH acknowledge support from the World Premier 
International Research Center Initiative (WPI Initiative), MEXT, Japan. RH and EK 
acknowledge support from the ChETEC COST Action (CA16117), supported by COST 
(European Cooperation in Science and Technology). AH acknowledges support from 
the National Science Foundation under Grant No. PHY-1430152 (JINA Center for the 
Evolution of the Elements) and from the Australian Research Council Centre of 
Excellence for All Sky Astrophysics in 3 Dimensions (ASTRO 3D), through project 
number CE170100013. JC acknowledges support by the Spanish Ministry of Economy, 
Industry and Competitiveness (MINECO) under grant AYA2017-83216-P.

\bibliography{biblio}

\end{document}